\documentstyle[fleqn]{revp}
\newfont{\sff}{cmssi12} 
\newfont{\bigsf}{cmss12 scaled 2000} 
\newfont{\midsf}{cmss12 scaled 1000} 
\newfont{\smlsf}{cmss12 scaled 600}
\newfont{\bigsff}{cmssi12 scaled 2000} 
\newfont{\sfi}{cmssi10} 

\newcommand{\pmonth}{January, 2000}
\newcommand{\spec}{Models and Theories of the Nuclear Mass, pp.~87-94}
\newcommand{\no}{26}

\begin{document}
\parindent 0pt
\parskip 12pt
\setcounter{page}{1}

\title{Pairing correlation involving the continuum states}

\author{Naoki Tajima\\
\sff
Department of Applied Physics, Fukui University}

\abst{
The Hartree-Fock-Bogoliubov equation for the ground states of
even-even atomic nuclei is solved using the canonical representation
in the coordinate space for zero range interactions like the Skyrme
force.
The gradient method is improved for faster convergence to the
solutions under constraint of orthogonality between canonical
orbitals.
Necessity of the cut-off of the pairing interaction is shown even when
the number of the canonical orbitals are restricted.
A repulsive dependence of the interaction on the pairing density is
introduced as an implementation of the cut-off which leaves the HFB
super matrix state-independent.
}

\maketitle
\thispagestyle{headings}

\section*{Introduction}

\def\baselineskipTaj{\baselineskip}
\newcommand{\refer}[1]{$^{\ref{bib:#1})}$}

In neutron-rich nuclei, the pairing correlation significantly involves
the continuum single-particle states.  This makes the HF+BCS
approximation inadequate due to the unlocalization of the neutron
density distribution and demands one to solve the
Hartree-Fock-Bogoliubov (HFB) equation without approximations.  The
solution in the coordinate space was first formulated in
Ref.~\refer{DFT84} using the quasi-particle states and performed for
spherically symmetric states. However, its application to deformed
states is difficult because there are quite a large number of
quasiparticle states even for a moderate size of
the normalization box (i.e., the cavity to confine the nucleons to
discretize the single-particle states).
Every HFB solution has an equivalent expression of BCS variational
form.
The single-particle states to construct the BCS type wavefunctions are
called the HFB canonical basis or sometimes the natural orbitals.
This expression was used to solve the HFB for spherical states
originally in Ref.~\refer{RBR97}.

Although spherical solutions can be obtained easily with present
computers (for zero-range forces),
deformed solutions are still difficult to obtain.  The
two-basis method$^{\ref{bib:GBD94},\ref{bib:THF96},\ref{bib:TFH97})}$
is the only one implemented so far for
neutron-rich nuclei.
Some recent developments like Ref.\refer{SNP98} are also in progress.

We have applied the canonical-basis method to deformed
states in a three dimensional cubic mesh representation with density
dependent delta interactions.\refer{Taj98a}
It has turned out to be a very efficient alternative approach to
obtain the solutions.  The origin of its effectiveness is that the
number of necessary single-\-particle basis states to describe the
ground state of a nucleus is proportional to the volume of the nucleus
in the canonical-basis method while it commensurates with the volume
of the normalization box in conventional methods.  The
difference of the number of the basis states amounts to a factor of $
10^1$ - $10^3$.

In this paper we discuss the canonical-basis formulation of the HFB,
the method to obtain the canonical-basis solutions, faster
gradient-method paths than a naive i\-mag\-i\-nar\-y-\-time evolution,
the necessity of the cut-off of the pairing interaction, and an
implementation of the cut-off in terms of an interaction dependent on
the pairing density.

\section*{HFB in the canonical representation}

To begin with, let us formulate the HF and the HFB in the
coordinate-space representation in order to elucidate a difficulty of
the HFB and to suggest its possible solution in terms of the
canonical-basis representation.  For the sake of simplicity, we
consider only one kind of nucleons and designate the number of
nucleons by $N$ in Eqs.~(\ref{eq:vac_hf})-(\ref{eq:dens_hfbn}), which
are in this section.  The $z$-component of the spin of a nucleon is
represented by $s$ (= $\pm \frac{1}{2}$).

In the HF, one should minimize $\langle \Psi | H | \Psi \rangle$
for single Slater-determinant states,
\begin{eqnarray}
\label{eq:vac_hf} 
|\Psi \rangle & = & \prod_{i=1}^{N} a^{\dagger}_i | 0 \rangle, \\
\label{eq:cre_op_hf} 
a^{\dagger}_i & = & \sum_{s} \int d \vec{r} \; \psi_{i} (\vec{r},s) \;
a^{\dagger}(\vec{r},s),
\end{eqnarray}
by varying $\{ \psi_i \} _{i=1,\cdots,N}$ under orthonormality
conditions $\langle \psi_i | \psi_j \rangle$ = $\delta_{ij}$.
The operator $a^{\dagger}_i$
creates a nucleon with a wavefunction $\psi_{i} (\vec{r},s)$.
The distribution function of the density of the nucleons
is related to the wavefunctions as
\begin{equation}
\label{eq:dens_hf} 
\rho ( \vec{r} ) = \sum_{s} \langle \Psi | a(\vec{r},s) a^{\dagger}(\vec{r},s)
| \Psi \rangle
= \sum_{i=1}^{N} \sum_{s} | \psi_i ( \vec{r},s ) |^2.
\end{equation}

In the HFB, the solution takes the following form,
\begin{eqnarray}
\label{eq:vac_hfbq} 
|\Psi \rangle & = & \prod_{i=1}^{I} b_i | 0 \rangle, \\
\nonumber     
b_i & = & \sum_{s} \int d \vec{r} \left\{
\phi_i (\vec{r},s) \; a(\vec{r},s) \right. \\
& & \left. + \varphi_i (\vec{r},s) \; a^{\dagger}(\vec{r},s) \right\},
\label{eq:ani_op_qp} 
\end{eqnarray}
where $b_i$ is the annihilation operator of a negative-energy
Bogoliubov quasi-particle with amplitudes
$\phi_i (\vec{r},s )$ for presence and
$\varphi_i ( \vec{r}, s )$ for absence.
$I$ is the number of the basis states of the employed
rep\-res\-enta\-tion.  For a three-dimensional Cartesian mesh
(3D-mesh) rep\-res\-enta\-tion,\refer{BFH85} it is the number of the
mesh points (times four when spin-orbit interactions are included) and
typically $10^4$-$10^5$.  One should vary
$\{ \phi_i, \varphi_i \}_{i=1,\cdots,I}$
under orthonormality conditions
\begin{eqnarray}
\nonumber          
& & \sum_{s}\int d \vec{r}
\left\{
    \phi_i^{\ast}   ( \vec{r}, s ) \phi_j    ( \vec{r}, s )
  + \varphi_i^{\ast}( \vec{r}, s ) \varphi_j ( \vec{r}, s )
\right\}  \\
\label{eq:norm_qp} 
& = & \delta_{ij}
\;\;\; ( 1 \le i \le j \le I),
\end{eqnarray}
and a constraint on the expectation value of the number of nucleons,
\begin{eqnarray}
\label{eq:num_hfbq} 
\langle \Psi | \hat{N} | \Psi \rangle & = &
\sum_{i=1}^{I} \sum_s \int d \vec{r} \; | \varphi_i (\vec{r},s) |^2 = N, \\
\label{eq:num_op} 
\hat{N} & = & \sum_s \int d \vec{r} a(\vec{r},s) a^{\dagger}(\vec{r},s).
\end{eqnarray}
The nucleon density for state (\ref{eq:vac_hfbq}) is expressed as
\begin{equation}
\label{eq:dens_hfbq} 
\rho ( \vec{r} ) = \sum_{i=1}^{I} \sum_{s}
| \varphi_i ( \vec{r},s ) |^2.
\end{equation}

The essential difference between the HF and the HFB is that one has to
consider only $N$ $\sim 10^2$ wavefunctions in the former while one
has to treat explicitly as many single-particle wavefunctions as the
number of the basis in the latter.

Owing to the Bloch-Messiah theorem,\refer{RS80}
the state (\ref{eq:vac_hfbq}) can be expressed
(for the ground states of even-even nuclei) as,
\begin{eqnarray}
\label{eq:vac_hfbn} 
|\Psi \rangle & = & \prod_{i=1}^{K}
\left( u_i + v_i \; a^{\dagger}_{i} \; a^{\dagger}_{\bar{\imath}} \right)
| 0 \rangle,\\
\label{eq:cre_op_hfbn} 
a^{\dagger}_i & = & \sum_{s} \int d \vec{r} \; \psi_{i} (\vec{r},s) \;
a^{\dagger}(\vec{r},s),
\end{eqnarray}
where $a^{\dagger}_{i}$ and $a^{\dagger}_{\bar{\imath}}$ create a
nucleon with wavefunction $\psi_i(\vec{r},s)$ and
$\psi_{\bar{\imath}}(\vec{r},s)$, respectively, which are called as
the canonical basis\refer{RS80} or the natural orbitals\refer{RBR97}
of the HFB vacuum $| \Psi \rangle$.
One must use $K=\frac{1}{2}I$ for the exact equivalence
between Eqs.~(\ref{eq:vac_hfbq}) and (\ref{eq:vac_hfbn}) in general cases.
When $| \Psi \rangle$ is a
time-reversal invariant state, which we assume in this paper, $\psi_i$
and $\psi_{\bar{\imath}}$ are the time-reversal state of each other.
In this case, only one wavefunction of each time reversal pair should
be counted as independent variables of the variational procedure.

To obtain solutions in the canonical-basis framework, one should vary
$\{ \psi_i , u_i , v_i \}_{i=1,\cdots,K}$
under three kinds of constraints, i.e, the orthonormality conditions,
\begin{eqnarray}
\nonumber 
\langle \psi_i | \psi_j \rangle & = &
\sum_s \int d \vec{r} \psi_i^{\ast} (\vec{r},s) \psi_j (\vec{r},s)\\
\label{eq:ortho_hfbn} 
& = & \delta_{ij} \;\;\;  ( 1 \leq i \leq j \leq K),
\end{eqnarray}
fixed expectation value of the number of nucleons,
\begin{equation}
\label{eq:num_hfbn} 
\langle \Psi | \hat{N} | \Psi \rangle = 2 \sum_{i=1}^K v_i^2 = N,
\end{equation}
and the normalization of the $u$-$v$ factors $u_i^2+v_i^2=1$.

The nucleon density is expressed as
\begin{equation}
\label{eq:dens_hfbn} 
\rho ( \vec{r} ) = 2 \sum_{i=1}^{K} \sum_{s} v_i^2
\left| \psi_i ( \vec{r},s ) \right|^2.
\end{equation}

Reinhard et al.\ regarded that the advantage of the representation
(\ref{eq:vac_hfbn}) over (\ref{eq:vac_hfbq}) is that one has to
consider only a single set of wavefunctions
$\{ \psi_i \}_{i=1,\cdots,I}$
unlike a double set
$\{\phi_i, \varphi_i \}_{i=\pm 1,\cdots,\pm I/2}$.\refer{RBR97}
However, we expect much greater benefit from the canonical-basis
representation. Namely, $i$ may be truncated as
$i \leq K$ = ${\cal O}(N)$ $\ll$ $\frac{1}{2}I$
to a very good approximation.  It is because $\psi_i$ appearing on the
right-hand side of Eq.~(\ref{eq:dens_hfbn}) must be a localized
function as $\rho(\vec{r})$ on the left-hand side, while the
orthogonality (\ref{eq:ortho_hfbn}) does not allow many low-energy
wavefunctions to exist in the vicinity of the nucleus.
For 3D-mesh representations, $I$ is proportional to the volume of the
cavity while $K$ is proportional to the volume of the nucleus. The
latter is $10^1$-$10^3$ times as small as the former.

Incidentally,
the situation is quite different in the quasiparticle formalism.
On the one hand, the localization of the density demands only the
localization of $\varphi_i$ through Eq.~(\ref{eq:dens_hfbq}) while
$\phi_i$ does not have to be localized in general.
On the other hand, the orthogonality condition (\ref{eq:norm_qp})
involves both $\varphi_i$ and $\phi_i$.
This discrepancy enables many quasiparticle
states having similar $\varphi_i$ to be orthogonal to each other by
differing their $\phi_i$.


\section*{Mean fields for zero-range interactions}

Let us present the effective Hamiltonian employed in this paper.
We adopt a density-dependent zero-range interaction.
Zero-rangeness makes the mean-field potentials local,
which is an essential advantage for coordinate-space solutions.
On the other hand, the omission of momentum dependences are merely for
the sake of simplicity and there will not be essential differences in
the formulation if we use the full-form Skyrme force.\refer{RS80}
Our force is expressed using the parameterization of the Skyrme force as,
\begin{eqnarray}
\nonumber 
& & \hat{v}(\vec{r}_1,s_1;\vec{r}_2,s_2)\\
\label{eq:mf_int} 
& = & \left( t_0 + \frac{1}{6} t_3
\rho\left( \vec{r}_1 \right)^{\alpha} \right)
\delta \left( \vec{r}_1 - \vec{r}_2 \right),
\end{eqnarray}
where $\vec{r}_i$ and $s_i$ are the position vector and the
spin state of the two interacting nucleons, $i=1,2$.
Dependence on the isospin state is redundant because of the zero-rangeness.
We adopt
$t_0$ = $-983.4$ MeV fm$^{3}$,
$t_3$ = $13106$ MeV fm$^{3+3 \alpha}$, and $\alpha$ = 0.98
when the force
is used to construct the mean-field (HF) potential.\refer{BV94}
We use different strengths to make the pairing potential. We express
the pairing force in the parameterization of Ref.\refer{TBF93} as
\begin{eqnarray}
& & \hat{v}_{\rm p}(\vec{r}_1,s_1;\vec{r}_2,s_2)
\nonumber \\        
& = & v_{\rm p}
\frac{1-P_{\sigma}}{2} \left( 1 - \frac{\rho (\vec{r}_1)}{\rho_{\rm c}} \right)
\delta \left( \vec{r}_1 - \vec{r}_2 \right).
\label{eq:pair_int} 
\end{eqnarray}
$P_{\sigma}$ is the exchange operator of spin variables.
$\frac{1}{2}(1-P_{\sigma})$ is a projector into spin-singlet states
so that the interaction acts only between like nucleons.
We use $\rho_{\rm c}$ = 0.32 fm$^{-3}$
(to roughly vanish the volume-changing effect\refer{TBF93})
and $v_{\rm p}=-440$ MeV fm$^3$.


When one considers both protons and neutrons, the
state of the nucleus is expressed as a product of two BCS type states
(\ref{eq:vac_hfbn}) for the protons and the neutrons:
\begin{equation}
\label{eq:vac_pn} 
| \Psi \rangle  = \prod_{\rm q=p}^{\rm n} \prod_{i=1}^{K}
\left( u_i + v_i \; a^{\dagger}_{i,\rm q} \; a^{\dagger}_{\bar{\imath},\rm q}
\right) | 0 \rangle,
\end{equation}
where q distinguishes between protons (p) and neutrons (n).
$a^{\dagger}_{i,{\rm p}}$ creates a proton having a wavefunction
$\psi_{i,{\rm p}}(\vec{r},s)$ while
$a^{\dagger}_{i,{\rm n}}$ creates a neutron with a wavefunction
$\psi_{i,{\rm n}}(\vec{r},s)$.
The product form is due to the pairing interaction
(\ref{eq:pair_int}) acting only between like nucleons.

For the sake of simplicity, we treat $N$=$Z$ nuclei without
Coulomb interaction in this paper.
In this case, the wavefunctions are the same between protons and neutrons,
i.e.,
$\psi_{i           ,{\rm p}}(\vec{r},s)$ =
$\psi_{i           ,{\rm n}}(\vec{r},s)$,
$\psi_{\bar{\imath},{\rm p}}(\vec{r},s)$ =
$\psi_{\bar{\imath},{\rm n}}(\vec{r},s)$.
Moreover, because the potentials are independent of the spin, the
wavefunctions $\psi_i(\vec{r},s)$ can be factorized into a product
of a spin wavefunction and a real function of the position,
which we write $\psi_i(\vec{r})$ in the following.
It holds
$\psi_{i,{\rm p}}(\vec{r})$ = $\psi_{\bar{\imath},{\rm p}}(\vec{r})$
= $\psi_{i,{\rm n}}(\vec{r})$ = $\psi_{\bar{\imath},{\rm n}}(\vec{r})$.

With the interactions (\ref{eq:mf_int}) and (\ref{eq:pair_int}), the
total energy for state (\ref{eq:vac_pn}) is written as,
\begin{eqnarray}
\label{eq:tot_eng} 
E & = & \langle \Psi | H | \Psi \rangle =
\int {\cal H}\left(\vec{r}\right) d \vec{r}, \\
\nonumber                      
{\cal H}(\vec{r}) & = &
\frac{\hbar^2}{2m} \tau(\vec{r})
+ \frac{3}{8}t_0 \rho(\vec{r})^2
+\frac{1}{16}t_3 \rho(\vec{r})^{2+\alpha} \\
& + & \frac{1}{8}v_{\rm p}
\left( 1 - \frac{\rho(\vec{r})}{\rho_{\rm c}} \right) \tilde{\rho}(\vec{r})^2,
\label{eq:ham_dens} 
\end{eqnarray}
where $m$ is the average of the proton and the neutron masses
divided by $1-1/A$ for the correction of the center-of-mass motion.
${\cal H}(\vec{r})$ is called the Hamiltonian density while function of
position $\vec{r}$ in the right-hand side are
\begin{eqnarray}
\label{eq:tau} 
\tau (\vec{r}) & =
g {\displaystyle \sum_{i=1}^{K}} v_i^2 | \vec{\nabla} \psi_i (\vec{r}) |^2
& \!\!\! \mbox{: kinetic energy density,} \nonumber \\
\label{eq:rho} 
\rho (\vec{r}) & =  g {\displaystyle \sum_{i=1}^{K}}
 v_i^2 | \psi_i (\vec{r}) |^2
&  \!\!\! \mbox{: density,} \nonumber \\
\label{eq:rho_tilde} 
\tilde{\rho}(\vec{r}) & = g {\displaystyle \sum_{i=1}^{K}}
 u_i v_i | \psi_i (\vec{r}) |^2
&  \!\!\! \mbox{: pairing density,} \nonumber
\end{eqnarray}
where $g=4$, which is
a factor to account for the situation that a wavefunction $\psi_i$
takes care of four nucleons for the spin and isospin degeneracy.

The mean-field potential $V$ and the pairing potential $\tilde{V}$
are defined as
\begin{eqnarray}
\label{eq:mf_pot} 
V & =
{\displaystyle \frac{\partial {\cal H}}{\partial \rho}} & =
\frac{3}{4}t_0 \rho +\frac{2+\alpha}{16}
t_3 \rho^{1+\alpha} -\frac{v_{\rm p}}{8 \rho_{\rm c}} \tilde{\rho}^2, \\
\label{eq:pair_pot} 
\tilde{V}
& = {\displaystyle \frac{\partial {\cal H}}{\partial \tilde{\rho}}}
& = \frac{1}{4}v_{\rm p} \left( 1 - \frac{\rho}{\rho_{\rm c}}
\right) \tilde{\rho},
\end{eqnarray}
in which $V$, $\tilde{V}$, ${\cal H}$, $\rho$, and $\tilde{\rho}$ are
local functions of $\vec{r}$ while $t_0$, $t_3$, $\alpha$, $v_{\rm
p}$, and $\rho_{\rm c}$ are constants.

The mean-field and the pairing Hamiltonians are
\begin{eqnarray}
\label{eq:ham_mf} 
h & = & - \frac{\hbar^2}{2m} \vec{\nabla}^2 + V, \\
\label{eq:ham_pair} 
\tilde{h} & = & \tilde{V}.
\end{eqnarray}
The quasiparticle states are the eigenvectors of the so-called
HFB super matrix composed of $h$ and $\tilde{h}$:
\begin{equation}
\label{eq:hfb_matrix} 
\left( \begin{array}{cc} h & \tilde{h} \\ \tilde{h} & h \end{array} \right)
\left( \begin{array}{c} \phi_\alpha \\ \psi_\alpha \end{array} \right)
= \epsilon_\alpha
\left( \begin{array}{c} \phi_\alpha \\ \psi_\alpha \end{array} \right).
\end{equation}
This is just an eigenvalue problem of a hermitian (because of
the time-reversal symmetry) matrix.
The canonical orbitals are also determined by $h$ and $\tilde{h}$ but in
a more complex way as described in the next section.

\section*{Gradient method for canonical-basis HFB}


In this section we describe a procedure to
obtain the canonical-basis solution of the HFB equation
directory, not by way of quasi-particle states.
Instead of minimizing $E$ =
$\langle \Psi | H | \Psi \rangle$ with $|\Psi\rangle$ given by
Eq.~(\ref{eq:vac_pn}) under constraints of Eqs.~(\ref{eq:ortho_hfbn})
and (\ref{eq:num_hfbn}), one may introduce a Routhian $R$,
\begin{eqnarray}
\nonumber           
R & = & E - \epsilon_{\rm F} \cdot g \sum_{i=1}^{K} v_i^2 \\
\label{eq:routhian} 
& - & g \sum_{i=1}^{K} \sum_{j=1}^{K} \lambda_{ij} \left\{
\langle \psi_i | \psi_j \rangle - \delta_{ij} \right\},
\end{eqnarray}
and minimize it without constraints.
$\epsilon_{\rm F}$ is probably the most familiar Lagrange multiplier,
whose physical meaning is the Fermi level.
In the definition (\ref{eq:routhian}), $K^2$ Lagrange
multipliers $\lambda_{ij}$ obeying hermiticity,
\begin{equation}
\label{eq:hermiticity} 
\lambda_{ij} = \lambda_{ji}^{\ast},
\end{equation}
are introduced instead of $\frac{1}{2}K(K+1)$ independent multipliers.
This hermitization of $\lambda$ is adopted in order to make $R$ real
so that two conditions, $\delta R / \delta \psi_i =0$ and $\delta R /
\delta \psi_i^{\ast} =0$, become equivalent and thus one has to
consider only one of them.
Note that $\delta_{ij}$ is subtracted from
$\langle \psi_i | \psi_j \rangle$,
in contrast to Ref.~\refer{RBR97},
which is in order to treat $\lambda_{ij}$ not as constants like
$\epsilon_{\rm F}$ but as functionals of the wavefunctions.

Using notations,
\begin{equation}
\label{eq:sp_eng_gap} 
\epsilon_i = \langle \psi_i | h | \psi_i \rangle, \;\;\;
\Delta_i = - \langle \psi_i | \tilde{h} | \psi_i \rangle,
\end{equation}
the stationary conditions of $R$ result in two kinds of equations.
One is $\partial R / \partial v_i =0$, which concerns the
occupation amplitudes $v_i$ and is fulfilled by
\begin{equation}
\label{eq:bcs} 
v_i^2 = \frac{1}{2} \pm \frac{1}{2} \frac{\epsilon_i -\epsilon_{\rm F} }
{\sqrt{(\epsilon_i- \epsilon_{\rm F} )^2+\Delta_i^2}}.
\end{equation}
The minimum of $R$ for the variations of $u_i$ and $v_i$ is obtained
(when $\Delta_i \ge 0$)
by taking the minus sign for the double sign in the right-hand side of
Eq.~(\ref{eq:bcs}) and
choosing the same sign as $v_i$
for $u_i$ = $\pm\sqrt{1-v_i^2}$.
The stationary condition for a wavefunction $\psi_i(\vec{r},s)$
gives the following equation:
\begin{eqnarray}
\nonumber 
& {\displaystyle \frac{1}{g} \frac{\delta R}{\delta \psi_i^{\ast}}} & =
{\cal H}_i \psi_i - \sum_{j=1}^{K} \lambda_{ij} \psi_j \\
\label{eq:gradient} 
& - &
\sum_{j=1}^{K} \sum_{k=1}^{K} \frac{\delta \lambda_{jk}}{\delta \psi_i^{\ast}}
\left\{ \langle \psi_j | \psi_k \rangle - \delta_{jk} \right\} = 0,
\end{eqnarray}
where
\begin{equation}
\label{eq:ham_i} 
{\cal H}_i = v_i^2 h + u_i v_i \tilde{h}.
\end{equation}
One can regard ${\cal H}_i$ as a state-dependent single-particle
Hamiltonian.  This dependence on states makes the orthogonalization
conditions essential to the method.  For HF,
the orthogonalization conditions are easily fulfilled because $\psi_i$
are eigenstates of the same hermite operator $h$ and thus are
orthogonal between themselves at the solution :
$\langle \psi_j | \psi_i \rangle \cdot (\epsilon_j-\epsilon_i)$ = 0.
The orthogonalization
procedure is needed only because states satisfying
the orthogonality are unstable for decaying into Pauli-forbidden
configurations.
On the other hand, for the canonical-basis HFB method, the
orthogonalization is essential because the single-particle
Hamiltonians ${\cal H}_i$ differs from state to state.  Therefore, the
determination of the explicit functional form of $\lambda_{ij}$ is the
most important part of the method.  Reinhard et al.\ have proposed
\begin{equation}
\label{eq:lambda1} 
\lambda_{ij} = \frac{1}{2}
\langle \psi_j | \left( {\cal H}_i + {\cal H}_j \right) | \psi_i \rangle.
\end{equation}
Let us reason on which grounds the above definition can be deduced.
Understanding these grounds is indispensable in order to modify the
definition later for faster convergences.
From the requirement that Eq.~(\ref{eq:gradient}) must
hold at the solution (where $\langle \psi_i | \psi_j \rangle =
\delta_{ij}$), one can deduce,
\begin{equation}
\label{eq:lambda2} 
\lambda_{ij} = \langle \psi_j | {\cal H}_i | \psi_i \rangle
\;\;\; \mbox{at the solution}.
\end{equation}
Eqs.~(\ref{eq:lambda1}) and (\ref{eq:lambda2}) are equivalent at the
solution because $\lambda_{ij}$ is defined to be hermite
by Eq.~(\ref{eq:hermiticity}).  Since this
hermiticity must hold at any points
to ensure the equality between the number of constraints and the
number of independent multipliers,
one should not adopt Eq.~(\ref{eq:lambda2}) but Eq.~(\ref{eq:lambda1}).


One can utilize the gradient method
to obtain the HFB solutions in the canonical-basis formalism.
The most naive implementation of the gradient method
agrees with
the imaginary-time evolution method in its first order approximation
of the size of the imaginary-time step $\Delta \tau$:
\begin{equation}
\label{eq:imag_evo} 
\psi_i \rightarrow
\psi_i - \frac{1}{g} \Delta \tau \frac{\delta R}{\delta \psi_i^{\ast}}.
\end{equation}
%


We have developed a 3D-mesh canonical-basis HFB program from scratch
according to the above formulation.
We take an example of our calculations using the program for the
ground state of $^{40}$Ca.
The wavefunctions are expressed with $39 \times 39 \times 39$ mesh
points with mesh spacing of $a$=0.8 fm.  We employed the 17-point
finite-difference approximation to the Laplacian.
Note that the requirement of
precision is higher for HFB than for HF because one has to treat
larger momentum components than the Fermi momentum in HFB.
The vanishing boundary conditions are imposed on the
boundary (the 0th and the 40th mesh points) and the wavefunctions are
anti-symmetrically reflected in the boundary to apply the
finite-difference formula.  We considered $K=20$ canonical basis,
which can contain $Kg=80$ (=$2 \times A$, $A=40$) nucleons.

For the imaginary time step size $\Delta \tau$, it must hold\refer{TTO96}
\begin{equation}
\label{eq:delta_tau} 
\Delta \tau \le \frac{2}{T_{\rm max}}, \;\;
T_{\rm max}=3 \; \frac{\hbar^2}{2m}
\left( \frac{\pi}{a} \right)^2.
\end{equation}
We took $\Delta \tau = 1/T_{\rm max}$.

In Fig.~\ref{fig:conv_naive}, the error of the second equality
of Eqs.~(\ref{eq:gradient}) neglecting the error of orthogonality,
i.e., ${\rm max}_{i=1,\cdots,K} |{\cal H}_i \psi_i - \sum_j
\lambda_{ij} \psi_j |$ is plotted with a solid line versus the number
of evolution steps.  The corresponding quantity for HF, ${\rm
max}_{i=1,\cdots,A/4} |h \psi_i - \langle \psi_i | h | \psi_i \rangle
\psi_i |$ , is also plotted with a dash line.  The figure demonstrates
that one can indeed obtain HFB solutions with the natural-orbital HFB
method in the 3D-mesh representation. We obtained similar convergence
curves for the error of the orthogonality and for the inconsistency
between the potential and the densities.

The speed of the convergence is, however, about ten times as slow as
the HF case. This is the subject of the next section.

\begin{figure}[h]
\begin{center}
\framebox{Figure \ref{fig:conv_naive}}
\end{center}
\caption{Convergence to the HF and HFB solutions for $^{40}$Ca.}
\label{fig:conv_naive}
\end{figure}

We adopted an additional procedure which is not indispensable to
obtain the solutions but effective to make the convergence more robust
and somewhat quicker:
After every 25th gradient-method steps, we orthogonalize
$\{ \psi_i \}$ with the Gram-Schmidt algorithm in the ascending order
of $\epsilon_i$, defined in Eqs.~(\ref{eq:sp_eng_gap}),
and then diagonalize the super matrix of the HFB
Hamiltonian (\ref{eq:hfb_matrix}) by expanding the quasi-particle
wavefunctions $(\phi_i,\varphi_i)$
($1 \le i \le K$ and their negative-energy partners)
in a $2K$-di\-men\-sional basis
$\{ \psi_i \}$ $\oplus $ $\{ \psi_i \}$,
and finally transform the resulting quasi-particle wavefunctions to
canonical orbitals and occupation amplitudes $\{ \psi_i, v_i\}$
to renew them.

Incidentally, the period of 25 steps may be too frequent because the
effect seems to saturate at periods around 100. We adopt the period of
25 in most calculations, however, because the increase in the
computation time is only a several percent of the total time with this
period.

\section*{Acceleration of the gradient method}


We show the origin of the slow convergence and present a solution of
the difficulty in this section.

Steepest-descent paths, which the gradient method draws,
depend on the choice of the independent variables.
For example,
Eq.~(\ref{eq:imag_evo}) is obtained when one uses
$({\rm Re} \; \psi_i, {\rm Im} \; \psi_i)$ as independent variables
to define the gradient vector and then express it in coordinates
$(\psi_i, \psi_i^{\ast})$.
If one uses scale-transformed wavefunctions $\chi_i \equiv
\alpha_i^{-1/2} \psi_i$, where $\alpha_i$ is a scaling factor,
a gradient step becomes,
\begin{equation}
\label{eq:grad_chi} 
\chi_i \rightarrow \chi_i - \frac{1}{g} \Delta \tau
\frac{\delta R}{\delta \chi_i^{\ast}},
\end{equation}
which is equivalent to
\begin{equation}
\label{eq:grad_psi} 
\psi_i \rightarrow \psi_i - \frac{1}{g} \alpha_i \Delta \tau
\frac{\delta R}{\delta \psi_i^{\ast}}.
\end{equation}
The change from Eq.~(\ref{eq:imag_evo}) to Eq.~(\ref{eq:grad_psi}) is
equivalent to multiplying $\alpha_i$ to the single-particle
Hamiltonian ${\cal H}_i$ in Eqs.~(\ref{eq:gradient}).

When one parameterizes the scaling factor as $\alpha_i$
= $v_i^{-2\nu}$, the modified single-particle Hamiltonian
becomes
\begin{eqnarray}
\alpha_i {\cal H}_i & = &
v_i^{2-2\nu} h + v_i^{1-2\nu} u_i \tilde{h}
\nonumber \\ 
& = & \left\{\begin{array}{rcrl}
v_i^2 h & + & v_i u_i \tilde{h} & (\nu=0),\\
v_i   h & + &     u_i \tilde{h} & (\nu=\frac{1}{2}),\\
      h & + & \frac{u_i}{v_i} \tilde{h} & (\nu=1).
\end{array}\right.
\label{eq:alpha_ham} 
\end{eqnarray}
When $\nu=0$ (i.e., $\alpha_i=1$), to which the imaginary-time evolution
(\ref{eq:imag_evo}) corresponds, the single-particle Hamiltonian
$\alpha_i {\cal H}_i$ (=${\cal H}_i$)
can be very small for canonical orbitals whose
$\epsilon_i$ is much higher than the Fermi level (i.e., $\epsilon_i$
$-$ $\epsilon_{\rm F}$ $\gg$ $\Delta_i$).  This smallness makes the
changes of such orbitals very slow.  On the other hand, for $\nu=1$,
the potential can be very deep for such high-lying orbitals
due to the factor $u_i/v_i$ in front of $\tilde{h}$.
In this case, however, the gradient step may be numerically dangerous.
We usually use $\frac{1}{2} \le \nu < 1$, which provides a fast and
numerically stable method of solution.


When one introduces the acceleration factors (i.e., $\alpha_i > 1$),
the multipliers $\lambda_{ij}$ should be modified from
Eq.~(\ref{eq:lambda1}) by the following reason.  In the computation
of a gradient vector given by the first of Eqs.~(\ref{eq:gradient}),
the last term takes much more computing time than the first two terms
due to $\delta \lambda_{jk} / \delta \psi_i^{\ast}$.  One can forget
the last term if the orthogonality relations (\ref{eq:ortho_hfbn}) are
fulfilled along the path of the steepest descent.  Let's suppose that
the relations are satisfied before a gradient-method step is taken and
require that they are conserved to the first order in $\Delta \tau$
after the step, i.e.,
\begin{equation}
\label{eq:orthog_path} 
\langle \psi_i' | \psi_j' \rangle = \delta_{ij} +{\cal O}
\left( \left( \Delta \tau \right)^2 \right)
\;\mbox{if}\;
\langle \psi_i | \psi_j \rangle = \delta_{ij},
\end{equation}
with
\begin{equation}
\label{eq:one_grad_step} 
\psi_i'= \psi_i - \alpha_i \Delta \tau
\left( {\cal H}_i \psi_i - \sum_j \lambda_{ij} \psi_j \right).
\end{equation}
Substituting Eq.~(\ref{eq:one_grad_step}) into Eq.~(\ref{eq:orthog_path})
and requiring the hermiticity (\ref{eq:hermiticity})
result in
\begin{equation}
\label{eq:lambda3} 
\lambda_{ij}= \frac{1}{\alpha_i+\alpha_j}
\langle \psi_j | \left( \alpha_i {\cal H}_i + \alpha_j {\cal H}_j \right)
| \psi_i \rangle.
\end{equation}

This form of $\lambda_{ij}$ fulfills the requirement that it should
agree with the expression (\ref{eq:lambda2}) at the solution as the
naive form of Eq.~(\ref{eq:lambda1}) does.
Two forms differ, however, before reaching the solution
if one assumes $\alpha_i \not= \alpha_j$ in general.
Therefore, in order to conserve the orthogonality to vanish the last
term in Eq.~(\ref{eq:gradient}) one must not use
Eq.~(\ref{eq:lambda1}) but Eq.~(\ref{eq:lambda3}).
We have indeed suffered from large errors of orthogonality by using
the naive form (\ref{eq:lambda1}).
On the other hand, by using the correct form (\ref{eq:lambda3}),
we have observed not only that the error does not grow during the evolution
but also that the error decreases without performing explicit
orthogonalization procedures periodically during the evolution.
This decrease should originate in the second order terms in $\Delta \tau$
in Eq.~(\ref{eq:orthog_path}), whose effects we did not consider.

\begin{figure}[h]
\begin{center}
\framebox{Figure \ref{fig:conv_accel}}
\end{center}
\caption{Convergence to the HFB solution for $^{32}$S.}
\label{fig:conv_accel}
\end{figure}

We compare the results of calculations between $\nu=0$ and
$\nu=\frac{1}{2}$ in Fig.~\ref{fig:conv_accel}.
The wavefunctions are expressed with $19
\times 19 \times 19$ mesh points with mesh spacing of $a$=1.0 fm.  We
considered $K=16$ canonical basis, which can contain $Kg=64$ (=$2
\times A$) nucleons.
The figure shows the convergence history to a HFB solution.
Four quantities are plotted as functions of the number of
gradient steps.  They are, from the top to the bottom, the total
energy $E$, the pairing gap $\bar{\Delta}$
(averaged with weight $u_i v_i$), the size
of quadrupole deformation $\beta$, and the size of triaxiality of
deformation $\gamma$. The last two quantities
are determined from the mass quadrupole moments.\refer{TTO96}

The dot curves were obtained without accelerations, i.e.,
with $\alpha_i = 1$ or $\nu=0$ in Eq.~(\ref{eq:alpha_ham}),
while the solid ones were obtained with the acceleration method
with $\alpha_i = 1/v_i$ or $\nu=\frac{1}{2}$.
One can see that the convergences of these quantities become
by far faster by using the acceleration method.
This result demonstrates that canonical-basis HFB can be solved
without very heavy numerical computations.

\section*{Cut-off of the pairing interaction for canonical-basis HFB}

Finally let us discuss on the cut-off schemes of the
paring interaction in relation to the canonical-\-basis HFB method.

Delta function forces without cut-off leads to a divergence of the
strength of the pairing correlation.\refer{TOT94}
In order to circumvent the divergence,
in the conventional method of solution, one usually takes only
quasiparticles whose excitation energy $\epsilon^{\rm qp}$ is lower
than some cut-off energy parameter $\epsilon_{\rm cut}$ to construct
the ground state.$^{\ref{bib:DFT84})}$
Namely Eq.~(\ref{eq:vac_hfbq}) is modified to
\begin{equation}
\label{eq:vac_hfbq_cut}
\vert \Psi \rangle = \prod_{i=1}^{I}
\theta \left( \epsilon_{\rm cut} - \epsilon^{\rm qp}_i \right) b_i |0\rangle
\end{equation}
where $\theta$ is the step function.

In the canonical-basis method, the restriction on the number of
canonical orbitals may prevent the divergence without introducing
explicitly a cut-off energy.
We have examined this idea by performing numerical calculations for
various situations.
Then, we noticed that observables sometimes jumps suddenly in the
course of long-time evolution.

\begin{figure}[h]
\begin{center}
\framebox{Figure \ref{fig:pair_div}}
\end{center}
\caption{Sudden changes in three
quantities during the course of a gradient-method evolution.
}
\label{fig:pair_div}
\end{figure}

An example is shown in Fig.~\ref{fig:pair_div}.
The calculation was done for a nucleus $^{32}$S on a
19$\times$19$\times$19 cubic mesh with a mesh spacing $a$=1 fm.
We considered $K=20$ canonical orbitals, to which we gave harmonic
oscillator wavefunctions at the beginning.
An 11-point formula was employed for the Laplacian.
The acceleration parameter in Eq.~(\ref{eq:alpha_ham}) is taken as
$\nu=0.7$.  At first we suspected that these jumps were due to the
acceleration method.  However, with smaller
$\nu$, we still observed jumps; only they come later.

We investigated the origin of the jumps and found that
each sudden change was due to a shrinkage of a high-lying (i.e.,
having large $\epsilon_i$) canonical orbital to a mesh point.  This
shrinkage can decrease the total energy of the nucleus for the
following reason: The contribution of a canonical orbital to the
kinetic energy density is proportional to its BCS occupation
probability $v_i^2$, while pairing density is proportional to $u_i
v_i$.  Because $v_i \ll 1$ and $u_i \doteq 1$ for high-lying orbitals,
it holds that $u_i v_i \gg v_i^2$. Therefore the increase in the
kinetic energy due to the shrinkage is easily compensated by the gain
of the pairing correlation energy at the shrunken point.

The observed jumps indicate that, in most cases (or maybe all the
cases), there are no potential barriers between such physically
meaningless solutions and the physically reasonable one.

Incidentally,
we confirmed that the lack of potential barriers was not due to the
discrete approximation of the Laplacian in the kinetic energy term:
The approximation was based on the Lagrange polynomial interpolation,
which tend to underestimate the expectation value for high momentum
components.  One might suspect that more accurate treatments of the
kinetic energy could restore a barrier between the physical and
unphysical solutions.  However, we observed similar jumps by using the
Fourier transformation with periodic boundary conditions,\refer{BH86}
which gives the exact result up to $\pi/a$.

We have decided that it is necessary to introduce cut-off for the
reliability of the method.

As the cut-off scheme,
we first employed cut-off factors which are dependent on orbitals.
In the BCS approximation, a smooth cut-off method\refer{BFH85}
is often utilized, in which the interaction is modified as
\begin{eqnarray}
\label{eq:cut_sen_force} 
\hat{v}_{\rm pair} & = & \!\! -G
\left( \sum_i \; f_i \; a^{\dagger}_i a^{\dagger}_{\bar{\imath}} \right)
\left( \sum_j \; f_j \; a_{\bar{\jmath}} a_j \right), \\
\label{eq:cut_func_e} 
f_i & = & f(\epsilon_i),
\end{eqnarray}
where $f(\epsilon)$ is a function of single-particle energy $\epsilon$.
The function takes on $\sim 1$ well below a
chosen cut-off energy and smoothly becomes zero above it.
In analogy to the smooth cut-off method,
we modified the pairing density as
\begin{equation}
\label{eq:cut_pair_dens} 
\tilde{\rho} = g \sum_{i=1}^{K} \; u_i v_i \; \vert \psi_i \vert^2
\rightarrow
g \sum_{i=1}^{K} \; f_i \; u_i v_i \; \vert \psi_i \vert^2
\end{equation}
with
\begin{equation}
\label{eq:cut_func_k} 
\begin{array}{l}
f_i = \exp \left( -\frac{\mu^2}{4} k_i^2 \right), \;\; \mu=1.2\mbox{fm},\\
 \\
k_i^2 = - \int \psi_i^{\ast} \triangle \psi d\vec{r}.
\end{array}
\end{equation}
We made the dependence to be on the kinetic energy ($\propto k_i^2$),
not on the mean-field energy $\epsilon_i$ as in Eq.~(\ref{eq:cut_func_e}),
to avoid a highly complicated expression of the gradient for the
latter case. (In BCS, such complications are just neglected.)

The result was successful to prevent the shrinkages to points.
However, this cut-off scheme has an disadvantage that the HFB super
matrix in Eq.~(\ref{eq:hfb_matrix}) cannot be defined.
It follows that we do not have well-defined quasiparticle states and
cannot utilize them to express the HFB ground state. This drawback is
rather serious because quasiparticles are useful to improve the
precision of HFB solutions obtained by the canonical-basis method and
to accelerate the convergence further, as described in the fourth
section.

As an alternative method, we have introduced a repulsive
pairing-\-density dependence to the pairing force,
Eq.~(\ref{eq:pair_int}),
in addition to the usual density dependence:
\begin{eqnarray}
& & \hat{v}_{\rm p}(\vec{r}_1,s_1;\vec{r}_2,s_2) =
v_{\rm p} \frac{1-P_{\sigma}}{2}
\nonumber \\ 
& \times &
\left\{  1
    - \frac{\rho (\vec{r}_1)}{\rho_{\rm c}}
    - \left(
          \frac{\tilde{\rho}(\vec{r}_1)}{\tilde{\rho}_{\rm c}}
      \right)^2
\right\}
\delta \left( \vec{r}_1 - \vec{r}_2 \right).
\label{eq:pair_int_m} 
\end{eqnarray}
A set of reasonable values of the parameters are
$v_{\rm p} = - 440$ MeV fm$^{3}$, $\rho_{\rm c} = 0.32$ fm$^{-3}$,
and $\tilde{\rho}_{\rm c} = 0.3$ fm$^{-3}$.

With forces (\ref{eq:mf_int}) and (\ref{eq:pair_int_m}), the expectation
value of the energy for $N=Z$ systems is expressed as a space integral
of a Hamiltonian density:
\begin{eqnarray}
{\cal H}(\vec{r}) & = &
\frac{\hbar^2}{2m} \tau (\vec{r}) + \frac{3}{8}t_0 \rho(\vec{r})^2
+\frac{1}{16}t_3 \rho(\vec{r})^{2+\alpha}
\nonumber \\ 
& + & \frac{1}{8}v_{\rm p}
\left\{ 1 - \frac{\rho(\vec{r})}{\rho_{\rm c}}
- \left( \frac{\tilde{\rho}(\vec{r})}{\tilde{\rho}_{\rm c}}\right)^2
\right\} \tilde{\rho}(\vec{r})^2.
\label{eq:ham_dens_m}
\end{eqnarray}
The mean-field potential $V$ remains the same as Eq.~(\ref{eq:mf_pot})
while the pairing potential $\tilde{V}$
has an additional term:
\begin{equation}
\label{eq:pair_pot_m} 
\tilde{V}
 = {\displaystyle \frac{\partial {\cal H}}{\partial \tilde{\rho}}}
 = \frac{1}{4}v_{\rm p} \left\{ 1 - \frac{\rho}{\rho_{\rm c}}
- 2 \left( \frac{\tilde{\rho}}{\tilde{\rho}_{\rm c}} \right)^2
\right\} \tilde{\rho}.
\end{equation}
With this new type of force, the shrinkage problem is completely
removed.

\begin{figure}[h]
\begin{center}
\framebox{Figure \ref{fig:rho_c_tilde}}
\end{center}
\caption{Effect of changing the pairing-density cut-off parameter
$\tilde{\rho}_{\rm c}$ (fm$^{-3}$)
on the pairing gap averaged with weight factor of $u_i v_i$
for $^{32}$S.}
\label{fig:rho_c_tilde}
\end{figure}

In Fig.~\ref{fig:rho_c_tilde}, we show the dependence of the pairing
gap (averaged with weight $u_i v_i$) on the parameter
$\tilde{\rho}_{\rm c}$, which controls the pairing density dependence.
The values are taken after 5,000 gradient steps.
The set up of the calculations are the same as in
Fig.~\ref{fig:pair_div} except for $\tilde{\rho}_{\rm c}$.
One can see that the pairing gap has
reasonable values with $\tilde{\rho}_{\rm c} \le 1$ fm$^{-3}$.
This is a good news because the pairing-\-den\-si\-ty-de\-pen\-dent
term, $-(\tilde{\rho}/\tilde{\rho_{\rm c}})^2$, can be small for the
values of $\tilde{\rho}$ which one finds in physical solutions, in
contrast to the usual density-dependent term, $-\rho/\rho_{\rm c}$,
which cancels roughly 50\% of the density-independent term inside
nucleus.  This situation is illustrated in
Fig.~\ref{fig:pair_dens_dep}. The plotted quantity is the
pairing Hamiltonian density $\tilde{{\cal H}}$, which is the
term in the second line of Eq.~(\ref{eq:ham_dens_m}).
The introduction of the new term
demands only little change of the other parameters of the force
if one adopts $\tilde{\rho}_{\rm c} = 0.3$ fm$^{-3}$.

\begin{figure}[h]
\begin{center}
\framebox{Figure \ref{fig:pair_dens_dep}}
\end{center}
\caption{Dependence of the pairing Hamiltonian density $\tilde{\cal{H}}$
on the pairing density $\tilde{\rho}$ for two values of a parameter
$\tilde{\rho}_{\rm c}$=0.3 fm$^{-3}$ (solid curve) and
$\tilde{\rho}_{\rm c}$=$\infty$ (dot curve).
The vertical scale is arbitrary.
These curves applies when $\rho \ll \rho_{\rm c}$.
}
\label{fig:pair_dens_dep}
\end{figure}

We show an example of the time evolution of $\epsilon_i$ in
Fig.~\ref{fig:sp_levels}.
The set up of the calculation is the same as in
Figs.~\ref{fig:pair_div} and \ref{fig:rho_c_tilde} except that a
7-point approximation is used for the Laplacian to favor the emergence
of unphysical solutions and make this calculation a very
severe test of the cut-off scheme.
The abscissa is the number of the gradient steps while the ordinate is the
expectation value $\epsilon_i$ of the mean-field Hamiltonian for all
the $K$ canonical orbitals.
The pairing density dependence parameter $\tilde{\rho}_{\rm c}$ is
set at
the standard value of 0.3 fm$^{-3}$ before step 3,000 (interval I)
and after step 6,000 (interval III).
Between steps 3,000 and 6,000 (interval II) $\tilde{\rho}_{\rm c}$
is increased temporally to 3 fm$^{-3}$, which results
in a too weak dependence to prevent the divergence.
One can see that an unphysical solution emerges without
the pairing-density dependence term (in interval II) and that
the unphysical solution
is suppressed (in interval I) and is quickly restored to a physical one
(in interval III) with the presence of the term.

Before ending the section, let us mention that our next subject
concerning the cut-off is whether repulsive momentum dependences of
the Skyrme force help to make the cut-off scheme more natural because
it has been known that such momentum dependences make well-\-developed
plateau before the divergence sets in.$^{\ref{bib:TOT94})}$

\begin{figure}[h]
\begin{center}
\framebox{Figure \ref{fig:sp_levels}}
\end{center}
\caption{Evolution history of $\epsilon_i$, i.e., the expectation values
of the mean-field Hamiltonian for canonical orbitals.
The pairing-density dependent term is switched on
in intervals I and III ($\tilde{\rho}_{\rm c}$ = 0.3 fm$^{-3}$).
It is suppressed to be very weak in interval II
($\tilde{\rho}_{\rm c}$ = 3 fm$^{-3}$).
}
\label{fig:sp_levels}
\end{figure}


\section*{Conclusions}

We have developed a method to obtain canonical-basis HFB solutions in
a coordinate-space three-dimensional (3D) Cartesian mesh
representation.  The features of our method are summarized as follows.

\noindent
i) It is not for spherical but for deformed nuclei and thus it can treat
both deformation and continuum pairing simultaneously.

\noindent
ii) It is not based on the oscillator-basis expansion but described in
the 3D Cartesian mesh representation and thus can treat e.g.\ deformed
halo-like orbitals.

\noindent
iii) There is a strong reason to believe that the necessary number of
canonical orbitals is much smaller than the number of single-particle
basis.

\noindent
iv) In order to perform variations under constraint of orthogonality
between the wavefunctions of the canonical orbitals, Lagrange
multipliers were introduced as functionals of the wavefunctions in
Ref.$^{\ref{bib:RBR97})}$.  We have clarified the necessary
conditions for the form of these Lagrange multiplier functionals.
We have modified the functionals appropriately so that they conserve
the orthogonality during the course of the accelerated evolutions
explained in the next item.

\noindent
v) We have found that the convergence to HFB solutions is very slow
when one employs a naive gradient method.
On the other hand, the convergence is quite rapid for Hartree-Fock
solutions, which neglects the pairing correlation.  We have
investigated the origin of the slow convergence and found that the
time scale of the gradient evolution is different from one canonical
orbital to another depending on their BCS occupation amplitudes $v_i$.
The difference ranges over many orders of magnitude.  We have
introduced an orbital-dependent acceleration method of the gradient
evolutions and could overcome the difficulty of the slow convergence.

\noindent
vi) We have examined the effects of the cut-off of the pairing
interaction.  The 3D mesh mean-field methods have a practical use only
for zero-range interactions like the Skyrme force presently. One needs
a cut-off for zero-range pairing forces, without which the pairing
correlation energy diverges to $-\infty.^{\ref{bib:TOT94})}$
We have found that zero-range forces need cut-off even when the number
of canonical orbitals are finite.  The divergence occurs through
shrinking of high-energy canonical orbital(s) to (a) mesh point(s).

\noindent
vii) To suppress the divergence, we have introduced a
pairing-\-density dependent interaction as a better choice than
orbital-dependent cut-off factors.


We believe that, by choosing a faster gradient path with the
acceleration method and adopting the cut-off scheme in terms of the
pairing-density dependence, the canonical-basis HFB method is now
fully understood and has the potential to become the standard method
to treat neutron-rich nuclei in HFB. The details of this work will be
published soon.\refer{Taj99b}


\newcounter{bean}
\begin{list}
{\arabic{bean})}{\usecounter{bean}
   \setlength{\labelwidth}{5mm}
   \setlength{\labelsep}{0.5mm}
   \setlength{\leftmargin}{5.5mm}
   \setlength{\rightmargin}{0mm}
   \setlength{\listparindent}{0mm}
   \setlength{\parsep}{0mm}
   \setlength{\itemsep}{0mm}
}
\baselineskip=\baselineskipTaj
\item \label{bib:DFT84} 
         J.~Dobaczewski, Flocard and Treiner:
         Nucl.\ Phys.\ {\bf A422}, 103 (1984).
\item \label{bib:RBR97} 
         P.-G.~Reinhard, M.~Bender, K.~Rutz, and J.A.~Maruhn:
         Z.\ Phys.\ {\bf A358} 277 (1997).
\item \label{bib:GBD94} 
         B.~Gall et al.: Z.\ Phys.\ {\bf A348}, 183 (1994).
\item \label{bib:THF96} 
         J.~Terasaki, Heenen, Flocard and Bonche:
         Nucl.\ Phys.\ {\bf A600}, 371 (1996).
\item \label{bib:TFH97} 
         J.~Terasaki, Flocard, Heenen and Bonche:
         Nucl.\ Phys.\ {\bf A621}, 706 (1997).
\item \label{bib:SNP98}
         M.V. Stoitsov, W. Nazarewicz, and S. Pittel:
         Phys.\ Rev.\ {\bf C58} 2092 (1998).
\item \label{bib:Taj98a} 
         N.~Tajima: proc.\ {\em Innovative Comp.\ Methods in Nucl.\
         Many-Body Problems}, Osaka (1997), World Scientific..
\item \label{bib:BFH85}  
         P.~Bonche, H.~Flocard, P.-H.~Heenen, S.J.~Krieger, and
         M.S.~Weiss: Nucl.\ Phys.\ {\bf A443}, 39 (1985).
\item \label{bib:RS80} 
         P.~Ring and P.~Schuck: {\em The nuclear many-body problem}
         (Springer, New York, 1980).
\item \label{bib:BV94} 
         F.L.~Braghin and D.~Vautherin: Phys.\ Lett.\ {\bf B333}, 289 (1994).
\item \label{bib:TBF93} 
         N.~Tajima, P.~Bonche, H.~Flocard, P.-H.~Heenen, and M.S.~Weiss:
         Nucl.\ Phys.\ {\bf A551}, 434 (1993).
\item \label{bib:TTO96} 
         N.~Tajima, S.~Takahara, and N.~Onishi:
         Nucl.\ Phys.\ {\bf A603}, 23 (1996).
\item \label{bib:TOT94} 
         S.~Takahara, N.~Onishi, and N.~Tajima,
         Phys.\ Lett.\ {\bf B331}, 261 (1994).
\item \label{bib:BH86} 
         D. Baye and P.-H. Heenen: J. Phys. {\bf A19} 2041 (1986).
\item \label{bib:Taj99b} 
         N.~Tajima: in preparation.
\end{list}

\vspace*{\baselineskip}
\fbox{
\fbox{
\begin{minipage}[t]{130mm}
This paper has been published in
RIKEN Review No.~26 (January, 2000), pp.~87-94.\\
The issue is devoted to the proceedings of an international symposium on
Models and Theories of the Nuclear Mass,
RIKEN, Wako-shi, Saitama, Japan, July 19-23, 1999.  \\
PDF files of all the issues of RIKEN Review can be obtained freely from
the following URL:
http://www.riken.go.jp/lab-www/library/publication/review/contents.html
\end{minipage}
}
}

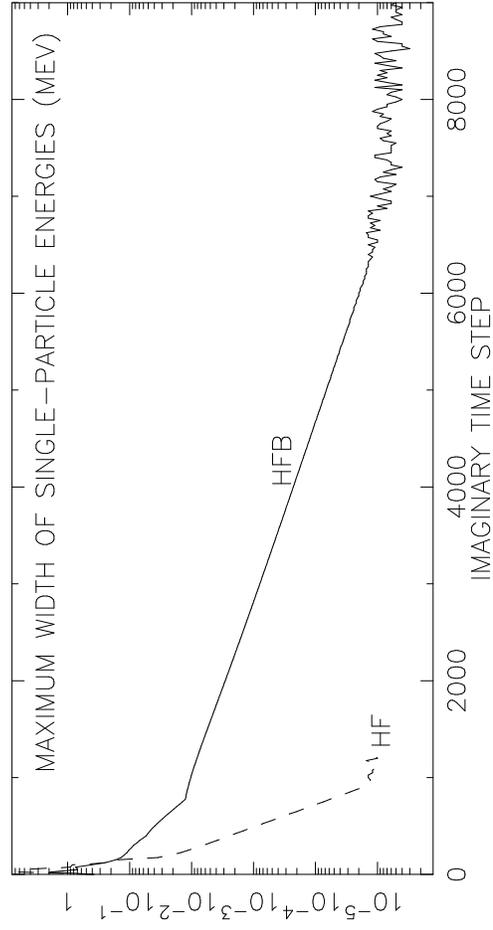
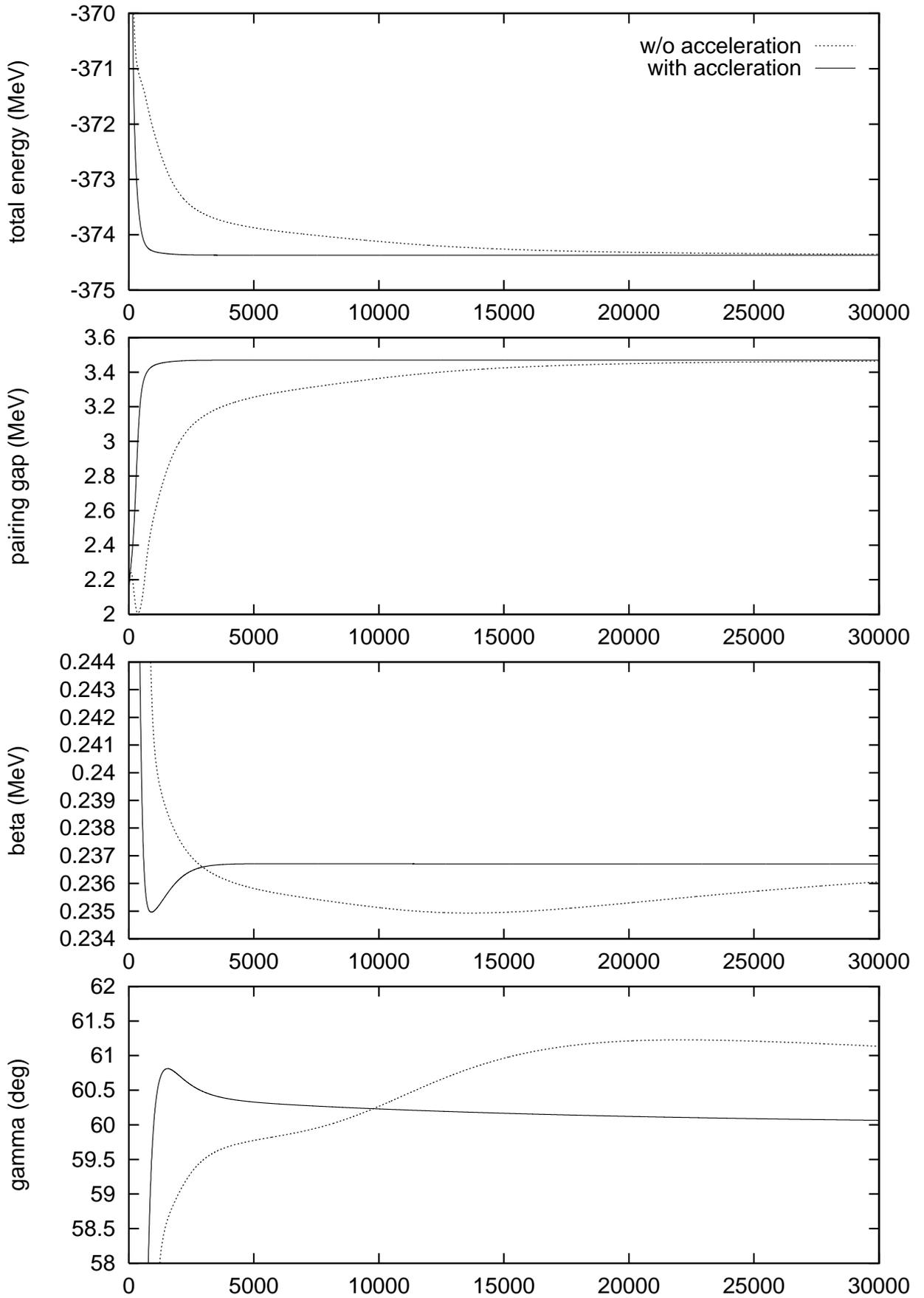
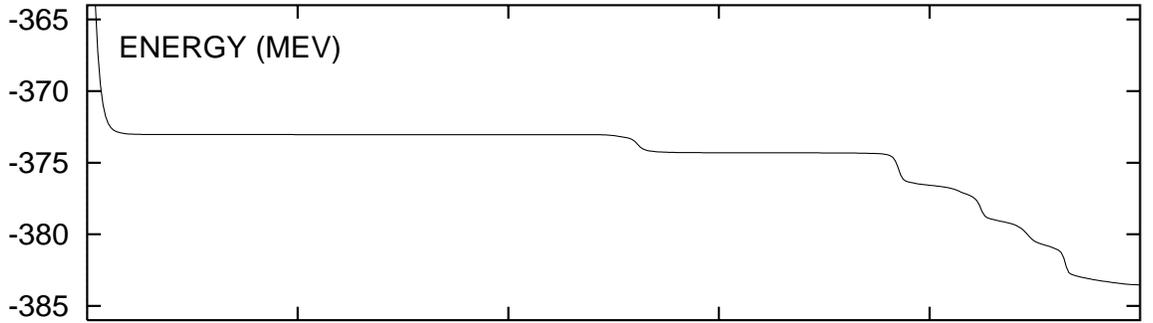
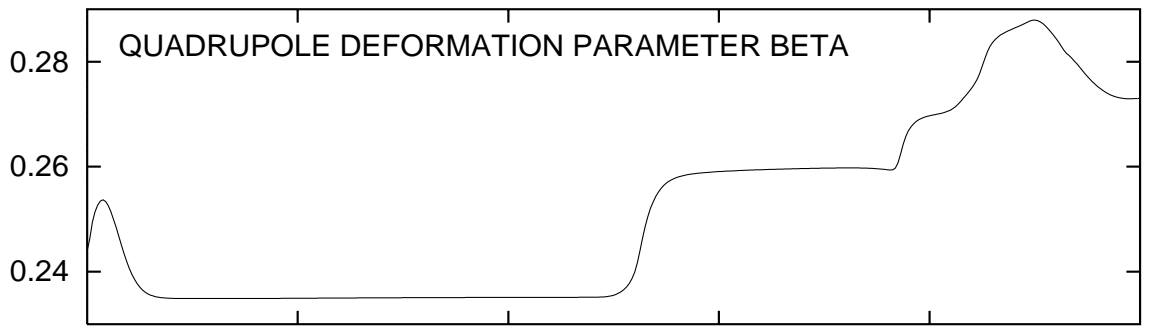
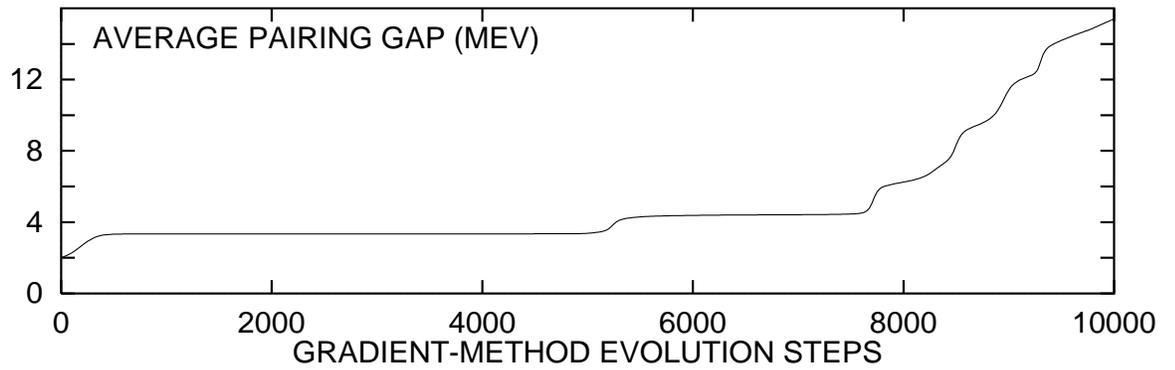
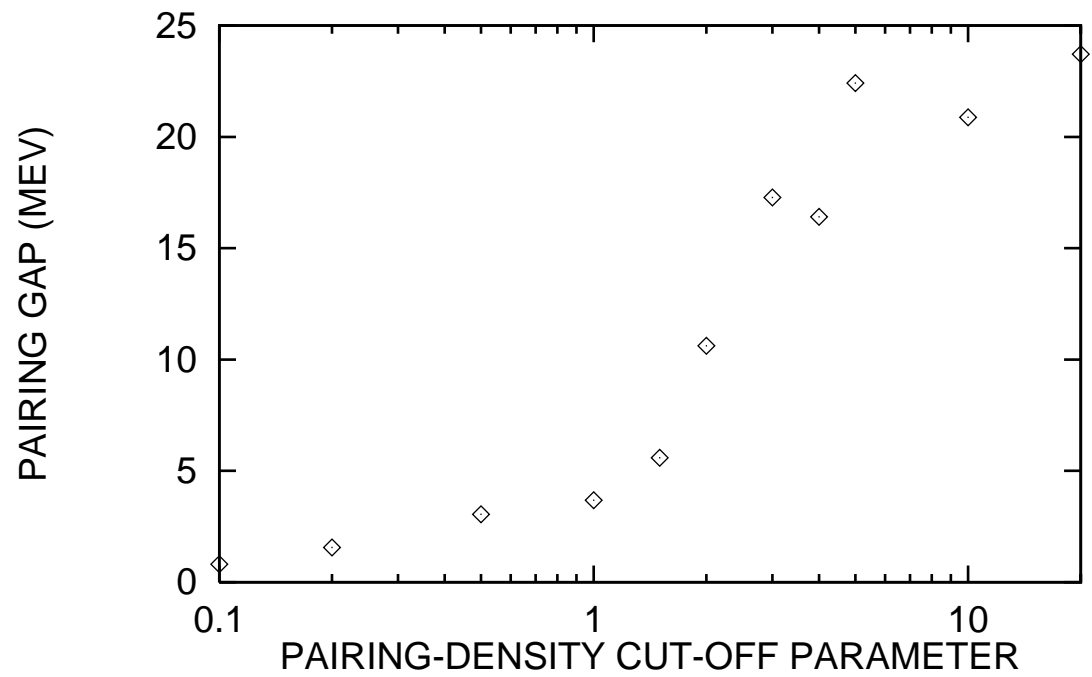
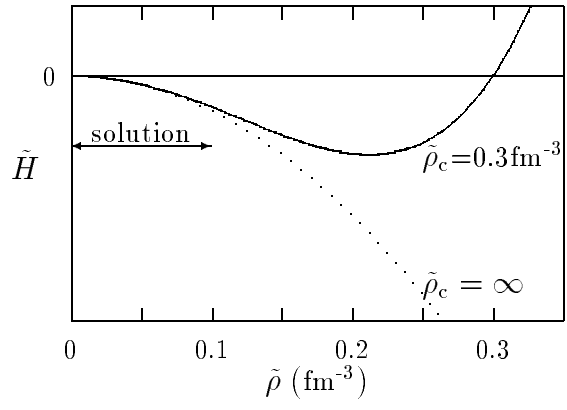
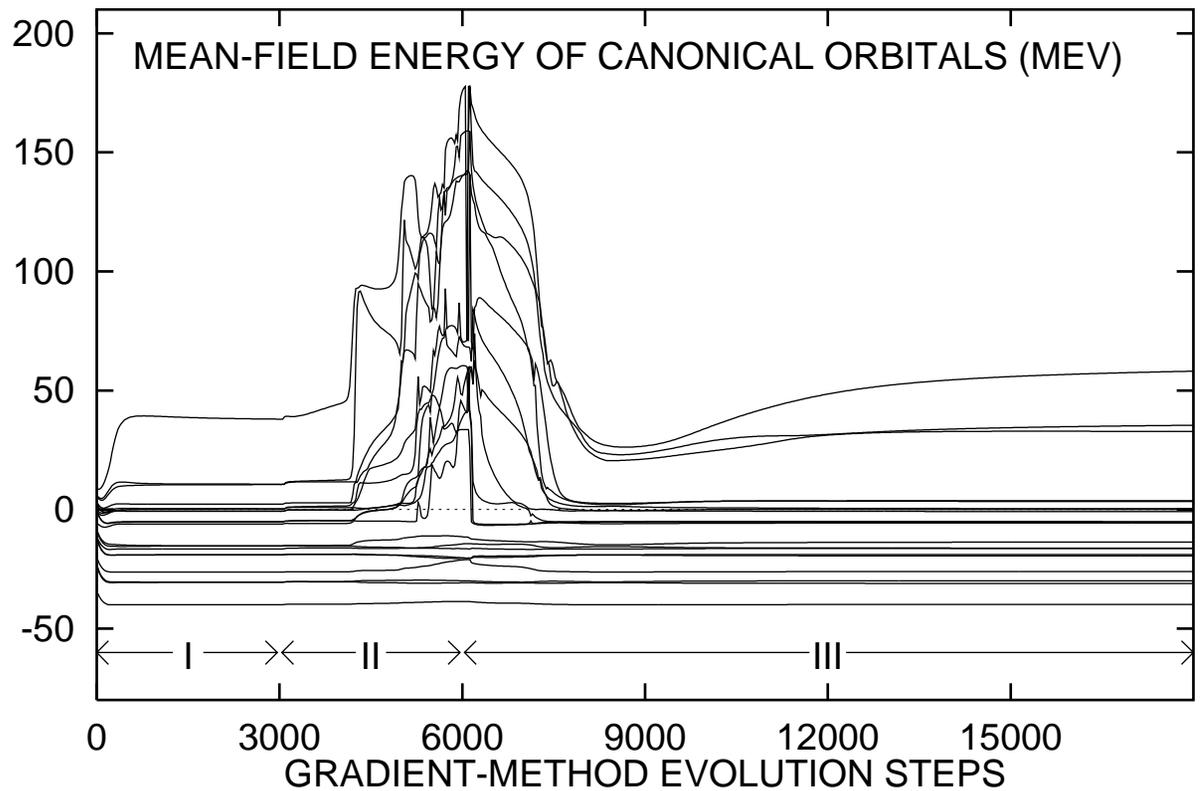
\end{document}